\documentclass[iop]{emulateapj}

\usepackage{graphicx}
\usepackage{natbib}
\usepackage{amssymb,txfonts}
\usepackage{multirow}
\usepackage{array}
\usepackage{sidecap}
\usepackage{hyperref}
\usepackage{epstopdf}
\usepackage{footnote}
\usepackage{tabularx}
\usepackage{booktabs}
\newcommand{\kms}{km~s$^{-1}$}
\newcommand{\degree}{\ensuremath{^\circ}}
\newcommand{\Rsun}{$R_\odot$}
\newcommand{\hinode}{\textit{Hinode}}
\newcommand{\yohkoh}{\textit{Yohkoh}}
\newcommand{\sdo}{\textit{SDO}}

\usepackage{longtable}

\shortauthors{Panesar et al.}

\begin{document}
 \title{HOMOLOGOUS JET-DRIVEN CORONAL MASS EJECTIONS FROM SOLAR ACTIVE REGION 12192}

\author{Navdeep K. Panesar\altaffilmark{1,2}, Alphonse C. Sterling\altaffilmark{1}, Ronald L. Moore\altaffilmark{1,2}}

\affil{{$^1$}Heliophysics and Planetary Science Office, ZP13, Marshall Space Flight Center, Huntsville, AL\\ 35812, USA}

\affil{{$^2$}Center for Space Plasma and Aeronomic Research (CSPAR), UAH, Huntsville, AL\\ 35805, USA}

\email{navdeep.k.panesar@nasa.gov}

\begin{abstract}
We report observations of homologous coronal jets and their coronal mass ejections (CMEs) observed by instruments
onboard the \textit{Solar Dynamics Observatory (SDO)} and \textit{Solar and Heliospheric Observatory (SOHO)} spacecraft.
The homologous jets originated from a location with emerging and canceling magnetic field at the southeast edge of the giant active region (AR)
of 2014 October, NOAA 12192. This AR produced in its interior many non-jet major flare eruptions
 (X and M class) that made no CME. During 20-27 October, in contrast to
 the major-flare eruptions in the interior, six of the homologous jets from the edge resulted in CMEs.
 Each jet-driven CME ($\sim$200-300 \kms) was slower-moving than most CMEs;
 had angular width (20\degree\ -- 50\degree) comparable to that
 of the  base of a coronal streamer straddling the AR; and was of the `streamer-puff' variety, whereby the preexisting streamer was transiently
 inflated but not destroyed by the passage of the CME.  Much of the transition-region-temperature
 plasma in the CME-producing jets escaped from the Sun, whereas relatively more of the transition-region
 plasma in non-CME-producing jets fell back to the solar surface. Also, the CME-producing jets tended
 to be faster and longer-lasting than the non-CME-producing jets. Our observations imply: each jet and CME resulted
 from reconnection opening of twisted field that
 erupted from the jet base; and the erupting field did not become a plasmoid as previously envisioned for streamer-puff CMEs, but instead the jet-guiding
 streamer-base loop was blown out by the loop's twist from the reconnection.

%  Our observations imply a
%  revision of a previous picture for the production of streamer-puff CMEs, where we now assume that the
%  jets result from small-scale filament eruptions. The CMEs result from helicity loading of the magnetic
%  arch into which the jet erupts. The twist-loaded arch then erupts to become a streamer-puff CME.
\end{abstract}

\section{Introduction}

Active region (AR) NOAA 12192 contained the largest sunspot group to date of solar cycle 24\footnote{http://www.thesuntoday.org/tag/sunspot/}.
The interior of this AR produced a multitude of big X-and M-class flares as well as
many B- and C-class flares during its passage across the solar disk from  2014 October 17 to 30.
All of these interior flares were confined, i.e. they did not produce CMEs
\citep{thalmann15,sun15,chen15}. The AR apparently produced a large fast CME on October 14, before it rotated onto the disk \citep{west15}.
During disk passage, it produced six `streamer-puff CMEs' \citep{bemporad05}, where the CME comes from a compact ejective eruption in a foot of
one loop of a coronal-streamer base magnetic arcade and the streamer transiently bulges out
but is not blown away completely by the passage of the CME \citep{moore07,jiang09}. Only one of these CMEs, accompanied by an M4.0 flare, was
previously reported \citep{thalmann15,chen15,lix15}. They all originated from the southeast edge of the AR, from
a series of coronal jets occurring at a neutral-line-containing subregion at that location, a neutral line separate
from that of the AR-interior confined flares.

``Jets" are dynamic, transient, collimated features that become long compared to their width.
Those with coronal emissions (`coronal jets') occur in coronal holes,
quiet regions, and ARs, and have a brightening at their base
\citep[e.g.][]{shibata92,shimojo96,sheeley99,cirtain07,nistico09}. 
Our observed jets might also be referred to as ``surges" \citep{zir88}. Jets are frequently described
as occurring on open-field regions (coronal holes, or at AR-coronal-hole
boundaries), whereas our jets here occur on relatively-large-scale closed loops of an
AR.

 There are various definitions of CMEs \citep[e.g.][]{sheeley09,vourlidas13}, here we use the term to mean a coronal
ejection that is listed in the LASCO CME 
catalog\footnote{http://cdaw.gsfc.nasa.gov/CME$\_$list}.

In addition to the CME-producing jets, many other jets from the same subregion did not
produce CMEs. Here we discuss the CMEs and the CME-producing jets, and differences between the CME-producing and the
non-CMEs-producing jets. We then present our interpretation that the CMEs were driven by magnetic
twist injected by the reconnection that made the CME-producing jets.

%%%%%%%%%%%%%%%%%%%%%%%%%%%%%%%%%%%%%%%%%%%%%%%%%%%%%%%%%%%%%%%%%%%%%%%%%%%%%%%%%%%%%%%%%%%%%%%%%%%%%%%%%%%%%%%%%%%%%%%%%%%

\begin{table*}
\begin{center}
\caption{Date and time for the observed jets, and their measured parameters. \label{tab:list}}
%\begin{tabular}{crrrrrrrrr}
\begin{tabular}{c*{9}{c}}
  \noalign{\smallskip}\tableline\tableline \noalign{\smallskip}
  \multicolumn{9}{l}
 {\textbf{a) CME-producing Jets:}} \\
  \noalign{\smallskip}\tableline \noalign{\smallskip}  
  Jet No & Date & Time\tablenotemark{a} &  Flare &  CME Speed\tablenotemark{b,}\tablenotemark{c}  &  CME  Angular &  Jet Speed\tablenotemark{d} & Jet Rise Dur. & Jet Width\tablenotemark{e}& Remote \\
        & (UT) &     & class & (\kms) & width (\degree) & (\kms) & ($\pm$ 5 min)  & ($\pm$ 1500 km) &  Bri.\& Dim.\\

\noalign{\smallskip}\tableline \noalign{\smallskip}

J1 & 20-Oct-14  & 18:43  & C6.2 &    187 & 40  & 190 $\pm$ 10  & 20 & 34000 &  Yes\\  
J2 & 22-Oct-14  & 16:52  & C5.8 &    281 & 20  & 310 $\pm$ 20 & 30 & 38000 & Yes \\  
J3 & 23-Oct-14  & 19:11  & C3.3 &    239 & 35  & 330 $\pm$ 20  & 50 & 26000 & No \\ 
J4 & 24-Oct-14  & 03:56  & C3.6 &    250 & 30  & 300 $\pm$ 20  & 45 & 34000 & Yes  \\ 
J5 & 24-Oct-14  & 07:37  & M4.0 &    677 & 50  & 400 $\pm$ 40 & 35 & 86000 &  Yes\\ 
J6 & 27-Oct-14  & 17:33  & M1.4  &   186 & 25  & ambiguous\tablenotemark{f}& - & - & -\tablenotemark{g} \\
\noalign{\smallskip}\tableline\tableline \noalign{\smallskip}
  %\multicolumn{9}{l}
  {\textbf{b) Non-CME-producing Jets:}}\\
 \noalign{\smallskip}\tableline \noalign{\smallskip}  
J8 & 22-Oct-14  & 02:31  & -    & -   & -  & 75 $\pm$ 10  & 35 & 19000 & - \\   
J9 & 22-Oct-14  & 05:51  & -    & -   & -  & 120 $\pm$ 20  & 10 & 15000 & - \\  
J10 & 22-Oct-14  & 10:46  & C1.9 & -   & -  & 140 $\pm$ 20 & 15 & 11000 & - \\  
J11 & 22-Oct-14  & 12:56  & -    & -   & -  & 50  $\pm$ 10 & 20 & 16500 & - \\  
J12 & 22-Oct-14  & 17:30  & C3.0 & -   & -  &  ambiguous\tablenotemark{h} & 10 & 13000 & - \\  
J13 & 22-Oct-14  & 20:11  & C3.0 & -   & -  & 150 $\pm$ 20 & 10 & 16000 & - \\  
J14 & 22-Oct-14  & 23:15  & C1.1 & -   & -  & 110 $\pm$ 10 & 25 & 13000 & - \\ 
\tableline
\end{tabular}

\footnotetext{ftp://ftp.ngdc.noaa.gov/STP/space-weather/solar-data/solar-features/solar-flares/x-rays/goes/2014/}
\footnotetext{http://cdaw.gsfc.nasa.gov/CME$\_$list}
\footnotetext{The uncertainty in the CMEs speed measurement is less than 10$\%$ \citep{yashiro04}.}
\footnotetext{The uncertainties are estimated from the time-distance plots.}
\footnotetext{Measured at a projected height of $\sim$72000 km from jet base.}
\footnotetext{This jet shows up well in the AIA 94 \AA\ images, but not in 304 \AA\ images. Due to its poor visibility in 304 \AA\ images, we were unable to follow the jet plasma well enough to measure its speed.}%%%The uncertainty in the jet speed and width measurements are  $\pm$ 20 \kms and $\pm$ 1500 km respectively.}
\footnotetext{AR was close to the west limb, obscuring any remote brightening/dimming.}
\footnotetext{Slower velocity (250 \kms) in the beginning, but faster ($>$650 \kms) later when a plug of plasma separates.}

\end{center}
\end{table*}

%%%%%%%%%%%%%%-----------------------------------------------------------------------
\section{Instrumentation and data}\label{data}

 In our analysis we used EUV images and movies from the
\textit{Solar Dynamics Observatory (SDO)}/\textit{Atmospheric Imaging Assembly (AIA)} to study the jets, and
we used images from the C2 coronograph of the \textit{Large Angle and Spectrometric Coronagraph}
(LASCO; \cite{brueckner95})  onboard the
\textit{Solar and Heliospheric Observatory (SOHO)} to study the CMEs.
 LASCO/C2  shows the outer corona between 2 \Rsun~and 6 \Rsun\ with a temporal
cadence of 12 minutes \citep{brueckner95}.

The \sdo/AIA images have a cadence of 12 s
and spatial resolution of 0\arcsec.6 pixel$^{-1}$ \citep{lem12}.
We used primarily 304 \AA\ and 193 \AA\ images\footnote{http://jsoc.stanford.edu/ajax/exportdata.html} to view
transition-region-temperature and coronal-temperature jet structures.
We derotated all the AIA images to a particular time and created movies with relatively coarse temporal cadence
(of 1-minute), which was sufficient for studying the jets' evolution.

The X-ray Telescope (XRT) on \hinode\ had coverage of only three
of our six CME-producing jets (J1, J2, and J5 in Table \ref{tab:list}). Each of these three jets was clearly visible in the XRT images.
So, probably all six of our CME-producing AIA jets were X-ray jets having cooler EUV components.

 We studied the photospheric magnetic field using \sdo/\textit{Helioseismic and Magnetic
Imager} (\textit{HMI}; \citealt{schou12})  line-of-sight magnetograms, which have cadence of 45 s and spatial resolution of
0\arcsec.5 pixel$^{-1}$ \citep{scherrer12}.%, observed in the \FeI\ 6173 \AA\ absorption line \citep{scherrer12}.

We found a total of six CME-producing homologous solar jets from AR NOAA 12192 between 2014 October 20 and 27.
These ejective jets and CMEs were first identified by looking at movies from JHelioviewer\footnote{http://www.jhelioviewer.org}.
Table \ref{tab:list}(a) lists the six jets and corresponding CMEs. %For each event, we downloaded the \sdo/AIA and LASCO C2 data.
We also studied the properties of seven non-CME-producing jets of 2014 October 22; see Table \ref{tab:list}(b).

\begin{figure*}
    \centering
     \includegraphics[width=\linewidth]{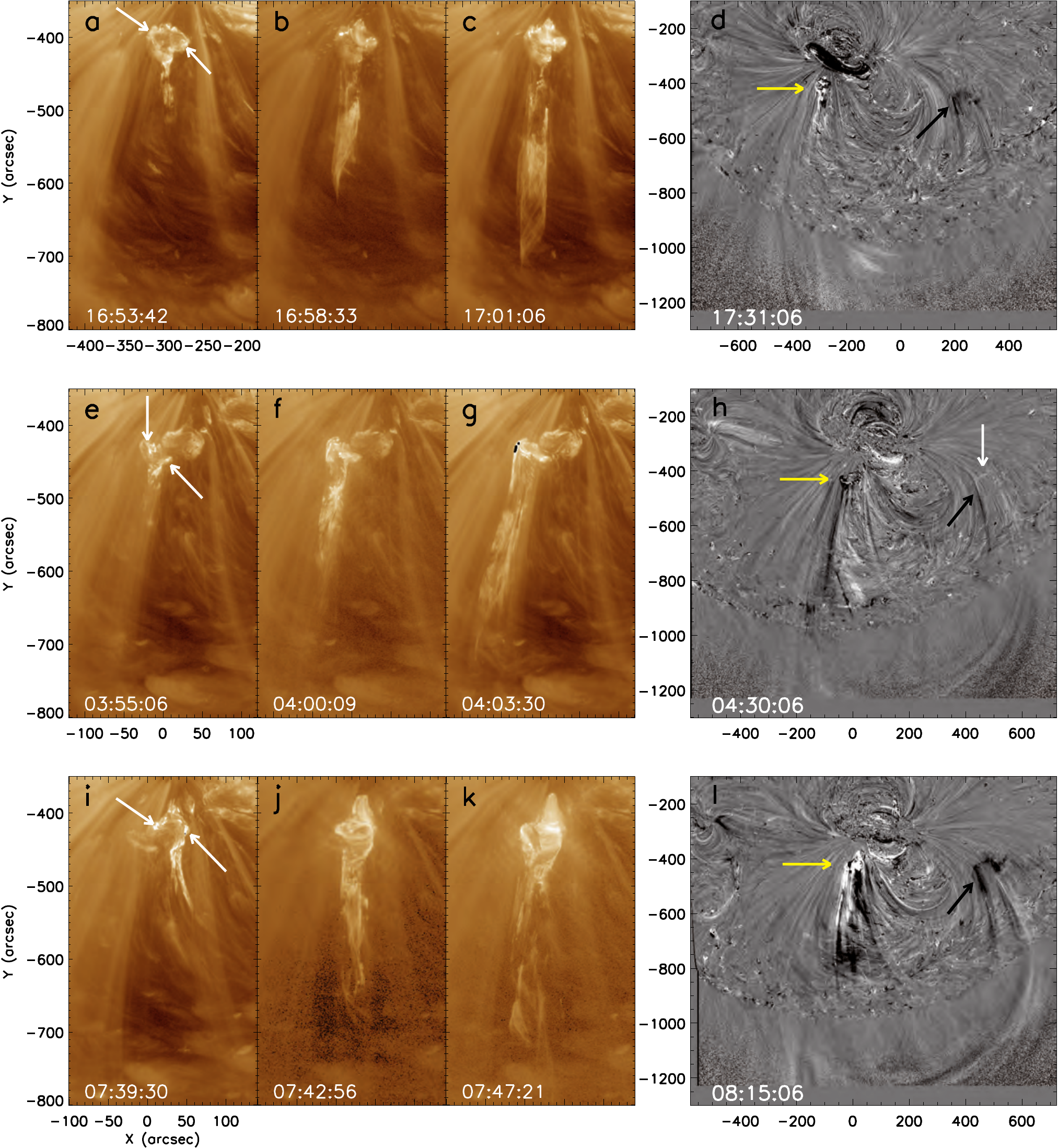}
      \caption{Evolution of homologous jets: AIA 193 \AA\ intensity images of J2 (a-c), J4 (d-f), and J5 (g-i)
      of Table \ref{tab:list}. %; J5 was the fastest, had the strongest accompanying flare, and produced the widest and fastest CME (Figure \ref{cme}(i)).
      In (a), (e), and (i), the arrows point to flare ribbons brightening in the jet base during the
      rise phase of the jet.  Panels (d), (h), and (l) show AIA 193 \AA\ base-difference images.
      The black and yellow arrows point to the remote dimmings and jet-origin region, respectively.
      The white arrow points to a remote  brightening.
      Animations of (d), (h), and (l) are available online; white arrows in selected frames show brightenigs.} \label{jet}
\end{figure*}

\vspace{0.2cm}

\section{Observations}\label{observations}

 Figure \ref{vl}(c) shows the AR. The highly-dynamic jetting location is on the south edge of the AR (Figure \ref{vl}(a)).
 This location produced six ejective eruptions, each of which produced a flare, jet and CME (Table \ref{tab:list}(a)). 
 These jets have characteristics of blowout jets \citep{moore10}. 
 These eruptions were reported in the \textit{Reuven Ramaty
 High-Energy Solar Spectroscopic Imager} \cite[\textit{RHESSI}\footnote{http://hesperia.gsfc.nasa.gov/hessidata/dbase/hessi$\_$flare$\_$list.txt};][]{lin02}
 flare list, as well as in the LASCO CME\footnote{http://cdaw.gsfc.nasa.gov/CME$\_$list} and NOAA flare 
 catalogs.\footnote{ftp://ftp.ngdc.noaa.gov/STP/space$-$weather/solar$-$data/solar$-$features/solar$-$flares/x$-$rays/goes/2014/goes$-$xray$-$report$\_$2014.txt}

\vspace{0.5cm}
\subsection{Evolution of Jets and CMEs}\label{evo}

Figure \ref{jet} shows three of the homologous jets observed by \sdo/AIA.
 Figure \ref{jet}(a-c) shows the progression of jet J2.
The white arrows in Figure \ref{jet}(a) point to brightenings in the base of
the jet as the jet begins to rise. Later (in Figure \ref{jet}(b)), the outward-moving jet spire extends higher in the corona
(Figures \ref{jet}(c) and \ref{vl}(c)).
The bright spire appears to extend along twisting magnetic field (e.g., at 16:58 UT in Figure \ref{jet}(b), also see MOVIE304).
 The transverse-motion in Figures \ref{vl}(d) and (f) (the insets) show definite twisting-motion tracks (traced by blue lines).
 The upper part of  (304 \AA) jet J2
leaves the AIA  field of view (FOV) at $\sim$17:21 UT (MOVIE304), showing that the material exceeded a height of 6.1 $\times$ 10$^5$ km
(which is the plane-of-sky distance from the jet base to the edge of the AIA FOV along the jet's path).
After that the lower part of the jet fades away slowly and
some of the jet material falls back to the solar surface ($\sim$18:26 UT). Figures \ref{jet}(e-g) and \ref{jet}(i-k) respectively show
 example images of jets J4 and J5.
  We find that the jets recur and  emanate from the same region, and have similar structure and development; that is, they are homologous \citep{dodson77}.

Figures \ref{jet}(d), (h), and (l) show the remote brightening and/or dimming at the other end of the loop during the jet eruptions.
%The brightenigs confirm that charged particles interact with the lower atmosphere during the reconnection. Similarly, {\bf dimming} shows the expulsion of the material during loop expansion.
We discuss the brightenings in Section \ref{discussion}. The dimmings support that the loop is ejected as the CME.
Such far-end brightenings and dimmings are not discernible in J3 (Table \ref{tab:list}), but that event is weaker than the others.

\begin{figure*}
    \centering
     \includegraphics[width=0.7\linewidth]{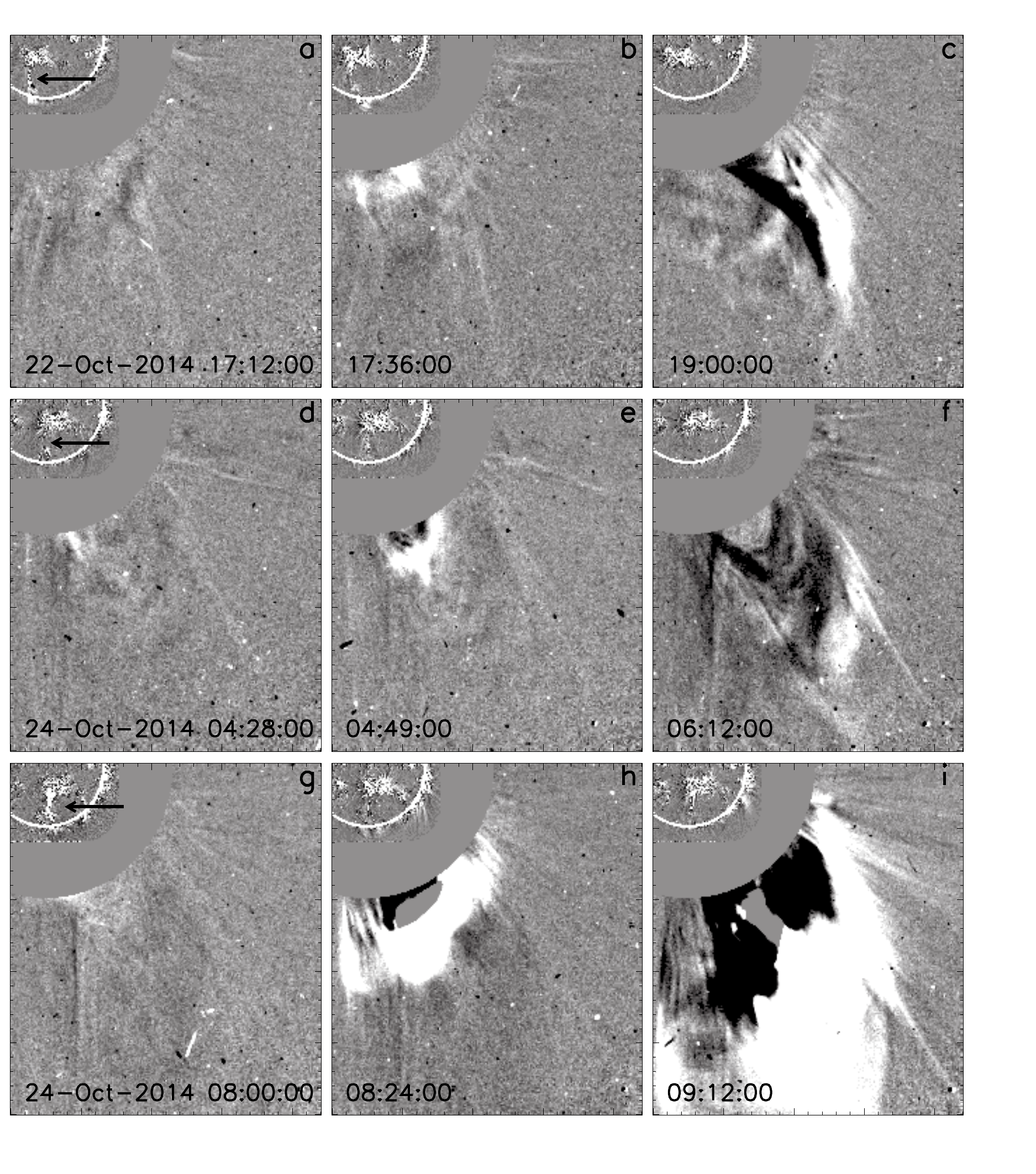}
      \caption{Progression of CMEs: (a-c), (d-f), and (g-i) are LASCO C2 running-difference images respectively showing the streamer-puff CMEs
      from jets J2, J4, and J5. In each frame, an \sdo/AIA 193 \AA\
      running-difference image is co-aligned with the C2 image. The outer edge of the AIA solar disk is outlined in white in
      each frame. The black arrows in (a), (d) and (g) point to the J2, J4 and J5 jets, respectively.
       Animation of J2 (a-c) is available online.} \label{cme}
\end{figure*}

In Figure \ref{cme}(a-c), we show the CME corresponding to jet J2 (Figure \ref{cme}(a) and MOVIECME). There is no indication of the CME (Figure \ref{cme}(a))
while the jet is still in the AIA FOV. But as soon as the jet
moves beyond the AIA FOV, the preexisting coronal streamer starts to inflate (see the non-difference LASCO
movie\footnote{http://lasco-www.nrl.navy.mil/daily$\_$mpg/2014$\_$10/}) as the CME is developing. %This behavior is the same for all the six jets.
Figure \ref{cme}(b) shows the early phase of the CME  when the jet was still escaping from the AIA FOV, and after that the CME continued to
expand and escape, and definitely shows twisted structure and untwisting motion (MOVIECME).
Our other streamer-puff CMEs show less definite evidence of twist in the LASCO C2 running-difference
movies\footnote{http://cdaw.gsfc.nasa.gov/CME$\_$list}.
The streamer was not blown out by the passage of the CME, only disturbed (i.e., transiently inflated).
Figures \ref{cme}(d-f) and \ref{cme}(g-i) respectively show the CMEs from jets J4 and J5.
These jet-driven CMEs had relatively narrow
angular width (20\degree - 50\degree; see Table \ref{tab:list}(a)), comparable to that of the streamer base ($\sim$40\degree).
The second CME observed on 2014 October 24 (from Jet J5) had the largest angular width (Figure \ref{cme}(i))
of the CMEs of Table \ref{tab:list}(a).

\subsection{Jet and CME Speeds}\label{speed}

\begin{figure*}
    \centering
     \includegraphics[width=0.9\linewidth]{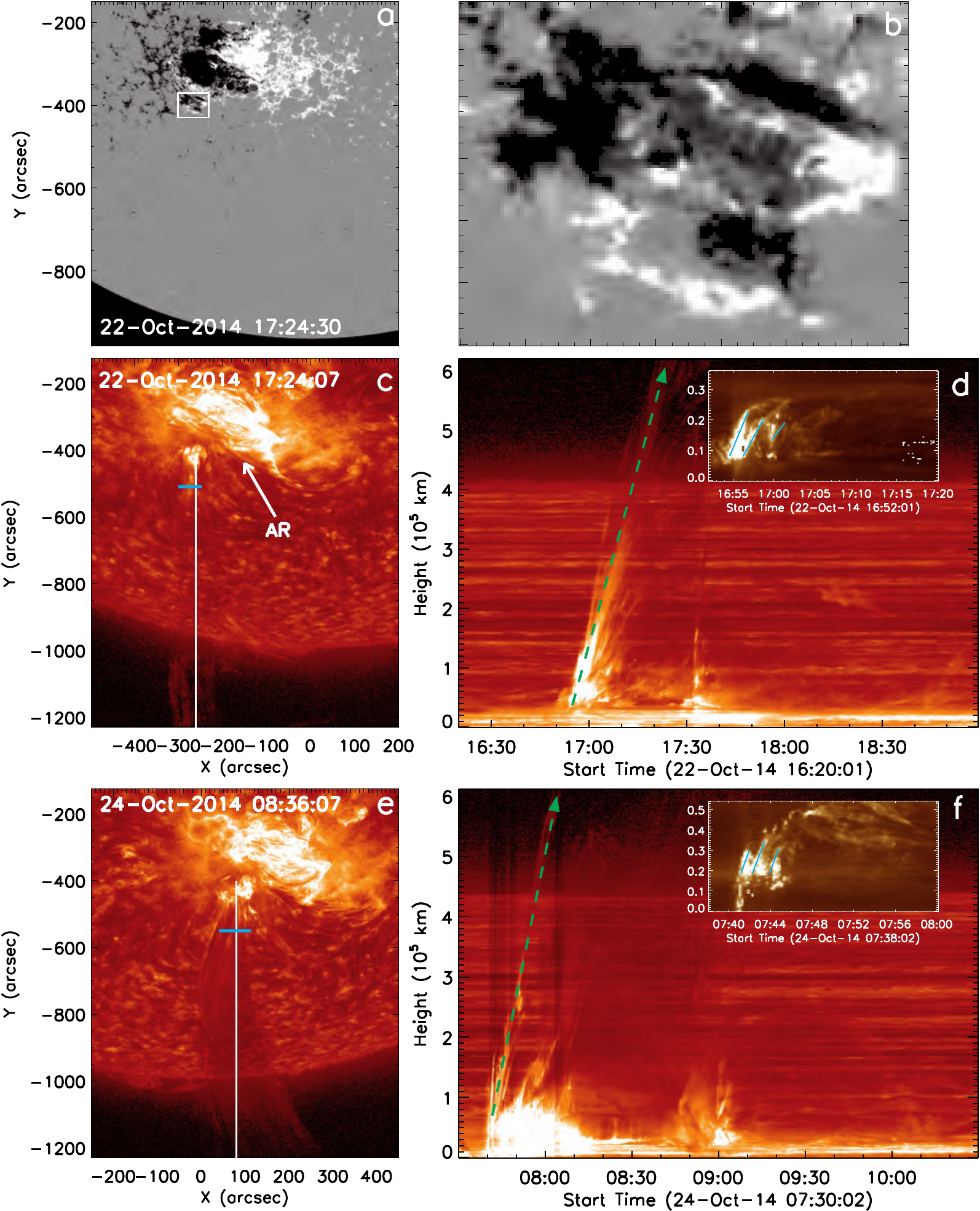}
      \caption{Jet outflow and spin: (a) HMI line-of-sight magnetogram of AR 12192 (b) The jet-producing region (white box of a).
       (c) AIA 304 \AA\ intensity image of jet J2, and (e) jet J5 of Table \ref{tab:list}. The white lines in (c) and (e)
      mark the positions of the time-distance plots in (d) and (f), respectively. Panels
      (d) and (f) show AIA 304 \AA\ intensity height-time-series images along the vertical lines in panel (c) and (e), respectively. Insets in
      (d) and (f) show the 193 \AA\ intensity time-series images along the blue lines in panel (c) and (e), respectively; they show 
      changes consistent with jet twisting with time. 
      Green dashed arrows in (d) and (f) are the paths used to calculate outflow speeds of the plasma.
      The x-axis of (a) is the same as (c). Animation of (c) is available online.} \label{vl}
\end{figure*}

All six CME-producing jets contain substantial transition-region (304 \AA) emitting material.
We measured the plane-of-sky speeds of the jets as they erupted and flowed outward (Table \ref{tab:list}).
 We take a straight line along the
main axis of the jet in 304 \AA\ (Figure \ref{vl}(c)) to construct a height-time plot. Figure \ref{vl}(d) shows the plot for
jet J2 along the white fiducial line in Figure \ref{vl}(c). It shows a bright outward flow of plasma starting at
$\sim$ 16:52 UT. The total duration of this jet is about 30 minutes. Plasma was propelled high into the corona
(Figure \ref{vl}(d) and MOVIE304); only a small fraction appears to fall back to the solar surface, while most
  material flowed out of the AIA FOV (similarly the other five jets of Table \ref{tab:list}(a)
largely left the AIA FOV). The slope of the green dashed arrow in Figure \ref{vl}(d) gives 310 \kms\ for the upflow speed of jet J2.

Figure \ref{vl}(d) shows
 another enhanced brightening beginning at 17:32 UT,
which is due to a subsequent jet that was not centered on the white line of Figure \ref{vl}(c).
This was a non-CME-producing jet (J12; Table \ref{tab:list}(b)). One can see  in MOVIE304
that plasma from this jet does not reach as high as the earlier jet (J2). The
material of jet J12 mainly becomes trapped in closed field lines. %This trapped material might subsequently leave the Sun slowly.

Figure \ref{vl}(e) shows jet J5, the largest jet of our data set, which erupts in conjunction with an M4.0 flare.
 Figure \ref{vl}(f) shows the height-time plot along the white fiducial line of
Figure \ref{vl}(e). This jet is much broader than other jets shown in Table \ref{tab:list}(a), and had a
  plane-of-sky
speed of about 400 \kms\ along the white line (slope of the green dashed arrow).
%This speed is much higher than for a quiet region jet ($\sim$60 \kms) reported by \cite{morton12}. %In the present study, the observed velocities are comparable with the results obtained by \cite{schmieder13}.
 This jet produced a CME with plane-of-the-sky speed of 680 \kms, which is twice that of the
CME from jet J2.
Speeds of all five jets and CMEs are listed in Table \ref{tab:list}(a).
The CME speeds are taken from the LASCO CME catalog. We extrapolated a linear fit  of the plane-of-the-sky speeds back
in time, and found the start times to match well with the jet start times for all six cases. This further supports that the CMEs originated from the jets.

To see the difference between the CME-producing jets and non-CME-producing jets, we measured the speeds,
durations and widths of both categories of jets (Table \ref{tab:list}(a,b)). Unlike the CME-producing jets,
the non-CME jets do not show a clear/visible and traceable ejection of transition-region plasma leaving the AIA FOV.
For calculating their outflow speeds, we tracked the leading edge of the jet in the 304 \AA\ images during
the jets' rise. (Using this method to measure speeds of the six CME-producing jets does not
result in values substantially different from those listed in Table \ref{tab:list}(a), which we found using the
time-distance plots such as in Figures \ref{vl}(d) and \ref{vl}(f). Thus our two speed-measurement methods are mutually-consistent.)
The non-CME jets that erupted in conjunction with substantial flares (J10, J12, J13 and J14) had
  higher speeds than the non-CME jets (J8, J9 and J11).

Figure \ref{vl}(a) shows the underlying magnetic field of the AR and jet region.
The jet base is a complex mix of emerging and canceling flux (Figure \ref{vl}(b)) throughout the disk passage of AR 12192.
 Therefore we can only conclude that
the jets occurred from a location where both flux emergence and cancellation were occurring.
 The properties of our jets are typical of those of \yohkoh/SXT-observed jets from ARs \citep{shimojo96}.

\section{Summary and Discussion}\label{discussion}

Our CME-producing jets have different properties than our non-CME-producing jets. We find the following: (a) relatively more of
the cool transition-region plasma
escaped from the Sun in the case of CME-producing jets whereas relatively more fell back to the
 surface in the non-CME-producing jets; (b) the CME-producing jets are faster (300 $\pm$ 75 \kms) than the
non-CME-producing jets (105 $\pm$ 40 \kms); (c) they tend to be longer in duration (mean duration and weighted
standard deviation are  35 and 10  minutes, respectively)
than the non-CME-producing jets (18 and 9 minutes, respectively); and d)
they are wider  (mean width and weighted standard deviation are  43,000 and 24,000 km, respectively)
than the non-CME-producing jets (15,000 and 27,00 km, respectively).
Our jet-driven CMEs are slower (speed $\sim$300 \kms) than average CMEs with flares \citep[$\geq$ 750 \kms, e.g.][]{sheeley99}.  %and narrower in  angular width  (20\degree\ - 50\degree)

\begin{figure}
    \centering
     \includegraphics[width=\linewidth]{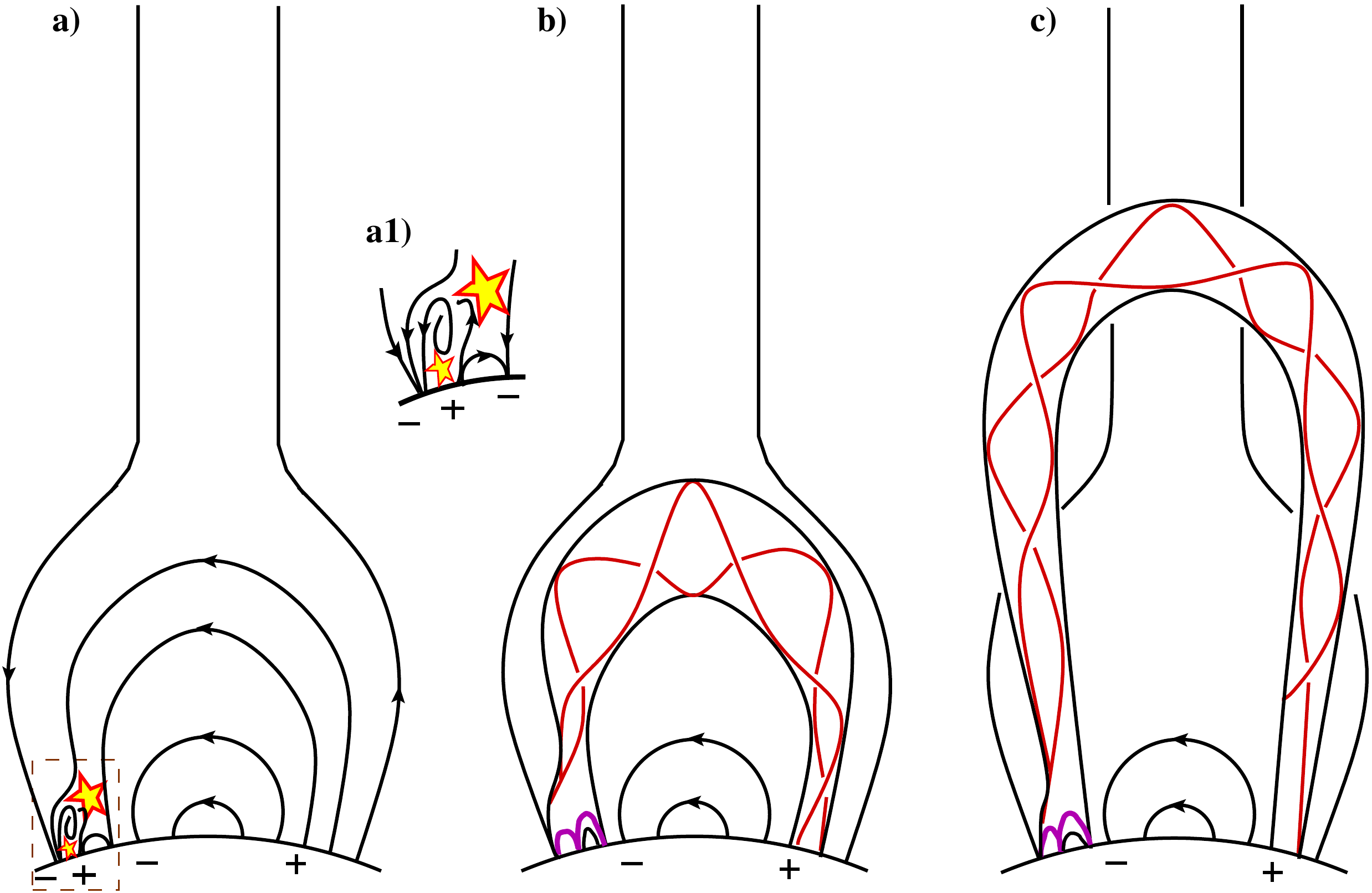}
      \caption{Schematic interpretation of the observations based on HMI, AIA, and LASCO images.
       These drawings depict the AR forming the helmet arcade below the streamer, viewed on the limb from the south.
      The helical black line in (a) represents twisted magnetic
      field in the jet base before and early during jet eruption.
      Stars show the locations where reconnection is taking place.
      Insert (a1) shows a zoomed-in view of the brown-boxed region of (a).
      The thick low magenta loops in (b) and (c) represent flare loops that result from internal (left) and external (right)
      reconnection of the erupting twisted field. (Complex flare loops at the jets' base in Figure \ref{jet} would correspond to the low-lying magenta
      loops of (b) and (c).)
      The red lines in (b) and (c) represent the twist transferred from the erupting field to the
      high-reaching jet-guiding coronal loop of the streamer-base helmet arcade by the
      external reconnection.
      The `+' and `--' labels are for positive and
      negative magnetic polarity, respectively.} \label{dia}
 \end{figure}

 Our CMEs result from jet eruptions. Recently, it has been found that jets in coronal holes are driven by minifilament eruptions \citep{sterling15}.
 The minifilaments reside in presumably-twisted magnetic field in the core of a small bipole between ambient open field and the
 minority-polarity side of a larger bipole.
  The minifilament-carrying bipole erupts and reconnects with the open field, producing the jet. It is plausible
 that our AR jets here operate the same way as \cite{sterling15} coronal hole jets \citep[cf.][]{lix15}.
  We further speculate that our jets lead to the CMEs as follows. During the reconnection, twist of the erupting-minifilament
 field is transferred to
 the newly-reconnected open field \citep{pariat09,torok09,shibata86,archontis13,fang14,moore15}. Since our AR jet eruptions occur in the
 foot of one loop of the streamer-base arcade, only that loop gets
 blown out rather than the whole arcade and whole streamer, and thus results in a streamer-puff CME \citep{bemporad05}. 
 
Figure \ref{dia} shows a schematic of this proposed process based on our observations. The
jet-producing region (dashed box in Figure \ref{dia}(a)) is embedded in the outskirt of the overall
arcade of loops of the AR and inside the arcade base of a LASCO-observed large streamer.
Based on \cite{sterling15}  and because there is cool transition-region plasma in our jets,
we assume that the jet-producing region (Figures \ref{dia}(a) and \ref{dia}(a1)) includes a sheared
field that contains a minifilament (\citealt{lix15} confirm that at least one of the jets of this region originates as an
erupting (mini)filament).  And since we observe spinning motion in our jets, we further assume that the minifilament
resides in twisted field \citep[e.g.][]{moore15}. Following the schematic of \cite{sterling15},
we envision that at the start of the jet the minifilament-holding field erupts, and undergoes two forms of magnetic reconnection:
(i) \textit {internal reconnection} among the legs of field inside of (i.e. \textit {internal} to) the erupting minifilament field
(lowest star in Figure \ref{dia}(a1)) that makes bright flare ribbons and loops at the jet base (shown as the jet-base left-hand-side
small magenta loop  in Figures \ref{dia}(b) and \ref{dia}(c)), and (ii) \textit {external reconnection} (highest star in Figure \ref{dia}(a1))
of the erupting minifilament field with a loop of the big arcade that is \textit {external} to the minifilament field. This external
reconnection: (1) makes the jet-base right-hand-side small magenta loop in Figures \ref{dia}(b) and \ref{dia}(c), and (2)
transfers twist from the erupting field to the reconnecting big loop (red twisted lines in Figure \ref{dia}(b)).
 We observed remote brightenings and dimmings at the far end of the erupting-CME loop (Figure \ref{jet}(d), (h), and (l) and Table \ref{tab:list}),
consistent with this picture; the brightenings are from high-speed electrons that are accelerated by the external reconnection,
escape along the big loop, and impact the far-end lower atmosphere \citep[e.g.][]{tang82}, and the dimmings are due to the big-loop blowout \citep[e.g.][]{moore07}. %Such brightenings were either weak or absent in J3.
During the eruptions, only one segment of the outer streamer arcade gets ejected
rather than the whole coronal streamer arcade because the jet eruptions occur  in the foot of only one loop of the arcade.
The added magnetic pressure from the added twist drives the arcade loop out to become the streamer-puff CME;
that is, the twist-loaded magnetic loop of the streamer base (Figure \ref{dia}(c)) erupts to
become the streamer-puff CME. After each eruption, the
opened field presumably recloses by reconnection, which allows the series of CMEs from the homologous jets \citep{sterling01,panesar15}.

The non-CME-producing jets are weaker,
and apparently do not transfer enough  twist to the streamer-base loop to blow it out as a discernible streamer-puff CME.

This picture of streamer-puff CMEs differs from that of \cite{bemporad05}. They proposed a scenario whereby
a flux-rope plasmoid explodes up the leg of the loop of the streamer-base arcade from a
compact ejective flare eruption, and explodes the loop top outward to become the streamer-puff CME \citep{bemporad05,moore07,jiang09}.
We propose that the CMEs are driven by the helicity loaded onto the magnetic-arch  loop from the erupting minifilament field as the jet forms,
 rather than by an erupting plasmoid as suggested by
\cite{bemporad05}.

  High-quality AIA data were not available for the events of \cite{bemporad05}.  We have since learned that many
 jets result from minfilament eruptions \citep{sterling15}, and so we suspect the jets we observe here occur that
 way also.  Moreover, AIA images of our events here show no indication of an erupting plasmoid (which would appear
 as a largely-intact closed-loop flux-rope structure; see Figures. 3(b) and 3(c) of \cite{bemporad05}).  Rather, apparently the minifilament
 field is entirely opened by the external reconnection and  ceases to be a plasmoid
 early in the jet-formation process (Figures \ref{dia}(a), \ref{dia}(a1) and \ref{dia}(b)),
 having become new big-arch field in the jet and new closed loops
 in the jet's base (Figure \ref{jet}(b, f, and j)).  Therefore, at least for the events presented here,
 the Figure \ref{dia} schematic explains the streamer-puff phenomenon better than does the schematic of \cite{bemporad05}.

\acknowledgments

This work was funded by the Heliophysics Division
of NASA's Science Mission Directorate through the Living With a Star Targeted Research and Technology Program, and by the \hinode\ Project. N.K.P. is supported by an appointment to the NASA Postdoctoral Program at the NASA MSFC, administered by USRA through a contract with NASA.

\bibliographystyle{apj}

\end{document}